# Ten Billion Years of Brightest Cluster Galaxy Alignments

Michael J. West[1], Roberto De Propris[2], Malcolm N. Bremer[3] & Steven Phillipps[3]

[1] Lowell Observatory, 1400 W Mars Hill Rd, Flagstaff, AZ 86001, USA

[2] Finnish Centre for Astronomy with ESO, University of Turku, Väisäläntie 20, 21500 Piikkiö, Finland

[3] Department of Physics and Astronomy, University of Bristol, Tyndall Avenue, Bristol BS8 1TL, UK

Corresponding author: mwest@lowell.edu

**A galaxy's orientation is one of its most basic observable properties. Astronomers once assumed that galaxies are randomly oriented in space, however it is now clear that some have preferred orientations with respect to their surroundings. Chief among these are giant elliptical galaxies found in the centers of rich galaxy clusters. Numerous studies have shown that the major axes of these galaxies often share the same orientation as the surrounding matter distribution on larger scales[1-6]. Using Hubble Space Telescope observations of 65 distant galaxy clusters, we show for the first time that similar alignments are seen at earlier epochs when the universe was only one-third its current age. These results suggest that the brightest galaxies in clusters are the product of a special formation history, one influenced by development of the cosmic web over billions of years.**

The most massive galaxies in the universe appear to know about their surroundings. It is well established that the major axes of brightest cluster galaxies (hereafter BCGs) are often elongated in the same direction as the galaxy cluster in which they reside and, furthermore, that clusters themselves are aligned with their neighbors, a remarkable coherence of structures over many millions of light years. With few exceptions[7-10], however, most studies of BCG alignments have focused on relatively nearby systems at redshifts $z < 0.1$. Examining more distant clusters can provide a glimpse into how these alignments have evolved over time, yielding insights into the processes that have shaped galaxies over the history of the universe.

With this motivation, we assembled a sample of 65 distant galaxy clusters with deep multi-color images available from the Hubble Space Telescope (HST) archive. These clusters, which are listed in Table 1, were discovered using a variety of techniques including optical and near-infrared imaging, x-ray detection, and the Sunyaev-Zel'dovich effect[11-16]. A few examples are shown in Figure 1. Although incomplete, this sample provides a representative selection of the most massive galaxy clusters at redshifts $0.19 < z < 1.8$, corresponding to look-back times ranging from two to ten billion years.

Because most galaxies in these HST fields lack spectroscopic redshifts, likely cluster members were identified based on their location along the 'red sequence,' a well-defined region in color-magnitude space occupied by passively evolving early-type galaxies. The number of red sequence galaxies detected ranges from more than one hundred in the richest low-redshift clusters in our sample to a dozen or so in the most distant clusters. Each galaxy's shape and orientation in the plane of the sky was measured from rest frame $r$-band HST images (for $z < 1$) or $Y$ and $JH$ images (for $z > 1$) by fitting a single-component Sérsic profile to the observed brightness distribution using the GALFIT software package (see Methods for more details).

The orientation of each cluster's principal axis was determined by computing the moments of inertia of the distribution of red sequence galaxies (see Methods for details), which are reliable tracers of the cluster mass distribution[17]. Cluster position angles are given in Table 1, along with 1σ uncertainties derived from bootstrap resampling, which are typically 10° to 20°. Thirteen clusters whose position angles were found to be uncertain by more than 25° were culled from the sample, leaving 52 clusters for subsequent analysis. Cluster orientations obtained from moments of inertia were found to be in good agreement with other independent determinations[18]. As a further check, for the clusters in the CLASH[12] sample with published mass models derived from gravitational lensing analysis we measured each cluster's principal axis and its orientation by fitting ellipses to the inferred mass distribution. The agreement is excellent in general, with a median difference of only 11° between the position angles obtained from moments of inertia versus gravitational lensing (see Methods for more details).

We first examine the general tendency for cluster galaxies of all luminosities to be aligned. Orientations were measured for 2137 individual galaxies in 22 CLASH clusters, reaching ~ 3-4 magnitudes fainter than the brightest member. We did not include all clusters for economy of effort, as the CLASH clusters are a homogeneous sample and contain most of the red sequence galaxies considered in this study. The acute angle, θ, between the position angle of each galaxy's major axis and that of its host cluster was computed and the results are shown in Figure 2. The uniform distribution between 0° and 90° is consistent with random orientations of galaxies in these clusters. This is confirmed by three different statistical tests for isotropy: Kuiper's V statistic, Rao's spacing test and the binomial test (see Methods for more details).

However, a very strong alignment tendency is seen when only the brightest member of each cluster is considered (Figure 3). The probability that the BCGs have random orientations with respect to their host clusters is very small, with $p$ = 0.000141, 0.000016 and 0.0152 according to the binomial, Kuiper V and Rao spacing tests, respectively. The second brightest and fainter galaxies show no significant alignment tendency, as is seen at lower redshifts[19,20]. Likewise, no correlation was found between alignments and a galaxy's absolute magnitude, surface brightness distribution as measured by Sersic index, ellipticity, nor on the magnitude difference between the first and second brightest cluster members. The primary factor that appears to determine whether a galaxy is aligned with its host cluster is that it must be the brightest cluster member, which suggests that there is something special about the birth and evolution of those galaxies. Other studies likewise support the view of BCGs as distinct from other cluster galaxies rather than just the statistical extreme of a single population[21-23], with alignments yet another piece of evidence that they are the product of a unique formation history.

Figure 4 shows the relative orientations of the brightest member galaxies with respect to their host clusters for the ten most distant clusters in our sample, all at $z > 1.3$. Despite the small sample size, there is clear evidence of BCG alignments at these epochs. The binomial test shows that the distribution seen in Figure 4 has a probability $p = 0.044$ of being consistent with random orientations, while the Kuiper V and Rao spacing tests indicate likelihoods of only $p = 0.019$ and $p = 0.051$, respectively. We conclude that the brightest galaxies in clusters have been aligned with their surroundings for at least the past ten billion years.

We emphasize that the alignments seen in Figure 3 and 4 are physical rather than a result of systematic errors. Cosmic shear is not expected to produce a false signal of intrinsic alignments within clusters, and any systematic errors that might arise in measuring galaxy orientations should not correlate with the distribution of galaxies on larger scales. In fact, given the uncertainties in the measured galaxy and cluster orientations, the intrinsic alignments of BCGs with their host clusters must be even stronger than seen in the figures. We note that the lack of alignments for non-BCGs is not an artifact of greater uncertainties in the position angles of fainter galaxies, because in many clusters the BCG and second-ranked member differ by only a few tenths of a magnitude in apparent brightness, and faint members of nearby clusters can appear brighter than the most luminous members of distant clusters.

There are several plausible theories for the origin of BCG alignments[24-26]. The most likely mechanisms are (a) anisotropic infall of matter into clusters along preferred directions (i.e., filaments) as seen in cosmological dark matter simulations, (b) primordial alignment with the surrounding matter distribution at the time of galaxy formation, (c) gravitational torques that gradually align galaxies with the local tidal field or (d) some hybrid of these. The results presented here do not allow us to differentiate between these scenarios except to note that the alignments must develop rapidly[27], as the highest redshift BCGs in our sample already have luminosities comparable to those of their low-redshift counterparts.

Numerical simulations could shed some light on the origin of BCG alignments. A growing number of studies have examined the expected alignment of galaxy- and cluster-size halos in the standard ΛCDM (Lambda Cold Dark Matter) cosmology, usually in the context of estimating the potential contamination of weak lensing measurements by intrinsic galaxy alignments[28-32]. Few, however, have specifically addressed the question of BCG alignments.

Because BCGs usually reside at or near the cluster center, this suggests that their alignment may be related to their special location. Indeed, it has been suggested that these galaxies could be viewed as 'proto-nuclei' of clusters[33]. Further study of BCG alignments at even higher redshifts could provide additional insights into the processes and timescales that have influenced the formation and evolution of these galaxies, the most massive in the universe, over billions of years.

**Table 1. Cluster sample**

| Cluster | Redshift $z$ | Major axis position angle* |
|---|---|---|
| Abell 383 | 0.189 | 16° ± 6° |
| Abell 209 | 0.209 | -69° ± 32° |
| Abell 1423 | 0.214 | 52° ± 6° |
| Abell 2261 | 0.224 | 35° ± 5° |
| RX J2129+0005 | 0.234 | 51° ± 13° |
| Abell 611 | 0.288 | 29° ± 8° |
| MS 2137 | 0.313 | 51° ± 25° |
| RX J1532+3020 | 0.345 | 51° ± 11° |
| RX J2248-4431 | 0.348 | 57° ± 12° |
| MACS J1932-2635 | 0.352 | 4° ± 15° |
| MACS J1115+0129 | 0.353 | 90° ± 3° |
| MACS J1720+3536 | 0.391 | 22° ± 2° |
| MACS J0416-2403 | 0.396 | 35° ± 4° |
| MACS 0429-0253 | 0.399 | 1° ± 16° |
| MACS 1206-0847 | 0.440 | 89° ± 12° |
| RX J1347-1145 | 0.451 | 56° ± 32° |
| MACS J1311-0310 | 0.494 | -82° ± 38° |
| MACS 0329-0211 | 0.450 | 20° ± 7° |
| MACS J1149.5+2223 | 0.544 | -37° ± 3° |
| MACS J1423+2404 | 0.545 | 42° ± 14° |
| MACS J0717+3745 | 0.548 | 62° ± 7° |
| MACS J2129-0741 | 0.570 | 34° ± 3° |
| SPT-CL J2331-5051 | 0.58 | -12 ± 13° |
| MACS J0647+7015 | 0.591 | 85° ± 15° |
| SPT-CL J0533-5005 | 0.60 | 41 ± 7° |
| SPT-CL J0559-5249 | 0.60 | -4 ± 11° |
| MACS J0744+3927 | 0.686 | -79° ± 7° |
| SPT-CL J0000-5748 | 0.70 | -10 ± 22° |
| SPT-CL J2337-5942 | 0.78 | 28 ± 7° |
| SPT-CL J2359-5009 | 0.78 | -51 ± 8° |
| SPT-CL J0102-4915 | 0.87 | -33 ± 2° |
| CL J1226+3332 | 0.89 | 83° ± 11° |
| SPT-CL J2040-5725 | 0.93 | 9° ± 14° |
| RX 1511 | 0.95 | -10° ± 44° |
| RCX 2319+0038 | 0.95 | 86° ± 10° |
| SPT-CL J0615-5746 | 0.97 | 68° ± 47° |
| XMMU J1229+0151 | 0.98 | 73° ± 42° |
| SPT-CL J2341-5119 | 1.00 | 87° ± 15° |
| SPT-CL J2342-5411 | 1.00 | 49° ± 19° |
| RCS J0221-0321 | 1.02 | -6° ± 9° |
| RCS J0220-0333 | 1.03 | 88° ± 13° |
| WARPS J1415+3612 | 1.03 | -22° ± 44° |
| RCS 2345-3632 | 1.04 | -28° ± 19° |
| SPT-CL J0546-5345 | 1.07 | 78° ± 26° |
| RDCS J0910+5422 | 1.11 | 31° ± 36° |
| SPT-CL J2106-5844 | 1.13 | 23° ± 25° |
| MOO J1142+1527 | 1.19 | 80° ± 12° |
| XLSS J0223-0436 | 1.22 | -64° ± 17° |
| RDCS J1252-2927 | 1.24 | 43° ± 15° |
| ISCS J1434+3427 | 1.24 | 44° ± 30° |
| MOO J1014+0038 | 1.24 | 65° ± 20 |
| ISCS 1429+3437 | 1.26 | 85° ± 21 |
| RDCS J0849+4452 | 1.27 | -26° ± 38 |
| SPT-CL J0205-5829 | 1.32 | 40° ± 20 |
| SpARCS-J0335 | 1.33 | -26° ± 8 |
| ISCS J1432.4+3436 | 1.35 | -39° ± 21 |
| ISCS J1434.5+3519 | 1.37 | 21° ± 9 |
| XMM J2235.3-2557 | 1.39 | 36° ± 14 |
| ISCS J1438+3414 | 1.41 | -42° ± 11 |
| XMMXCS J2215.9-1738 | 1.47 | 45° ± 35 |
| SPT-CL J2040-4451 | 1.48 | -82° ± 16 |
| XDCP J0044-2033 | 1.59 | -52° ± 49 |
| SpARCS-J0330 | 1.63 | 43° ± 12 |
| IDCS J1426.5+3508 | 1.75 | -66° ± 11 |
| JKCS 041 | 1.80 | -80° ± 18 |

\* Position angles are measured north through east. Uncertainties are estimated from bootstrap resampling.

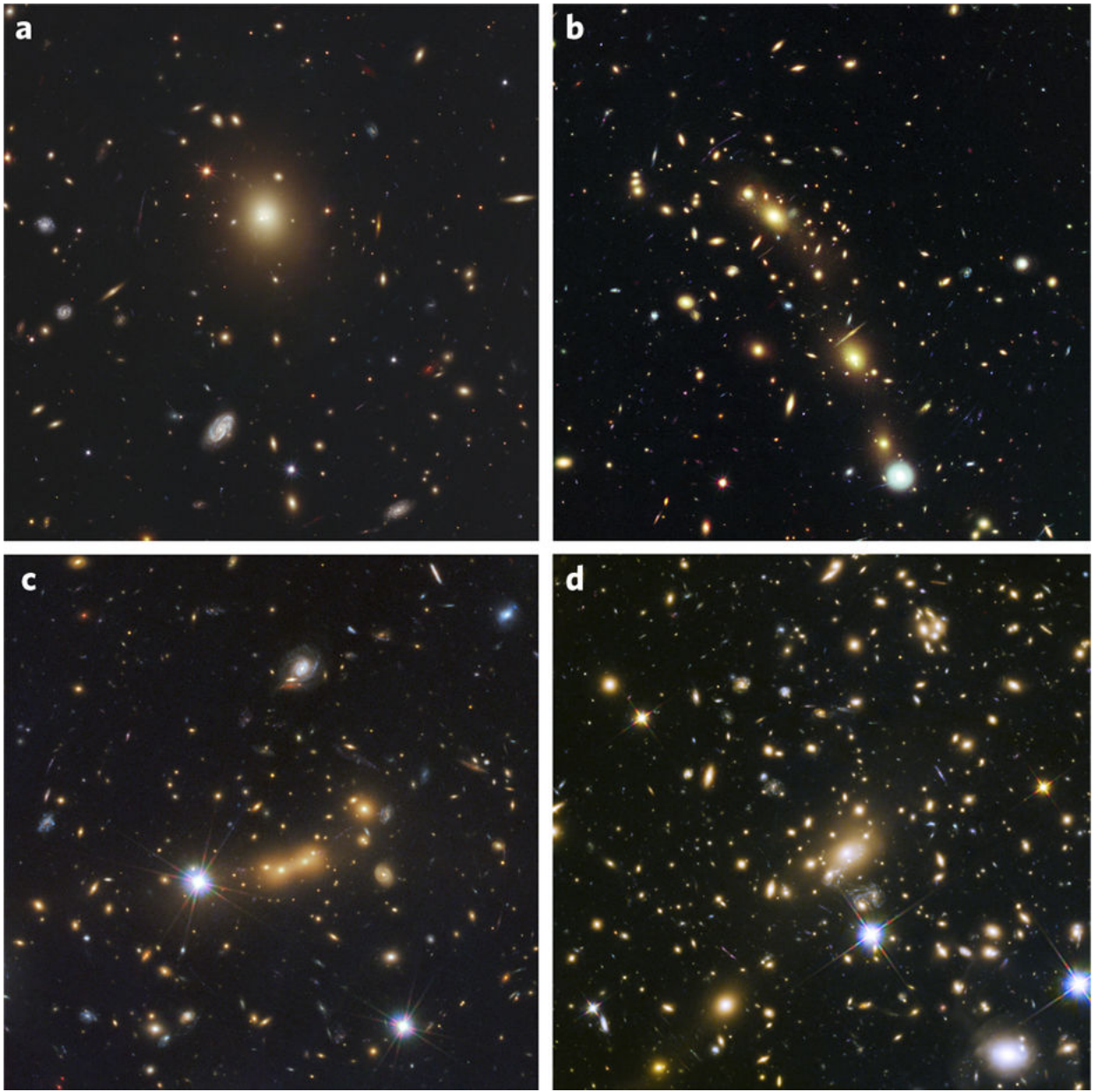

Figure 1. Hubble Space Telescope images of four distant galaxy clusters: a) Abell 2261 ($z$ = 0.224), b) MACS J0416.1−2403 ($z$ = 0.396), c) MACS J0647.7+7015 ($z$ = 0.591), and d) MACS 1149.5+2223 ($z$ = 0.544). Image credit: NASA, ESA, M. Postman and the CLASH team.

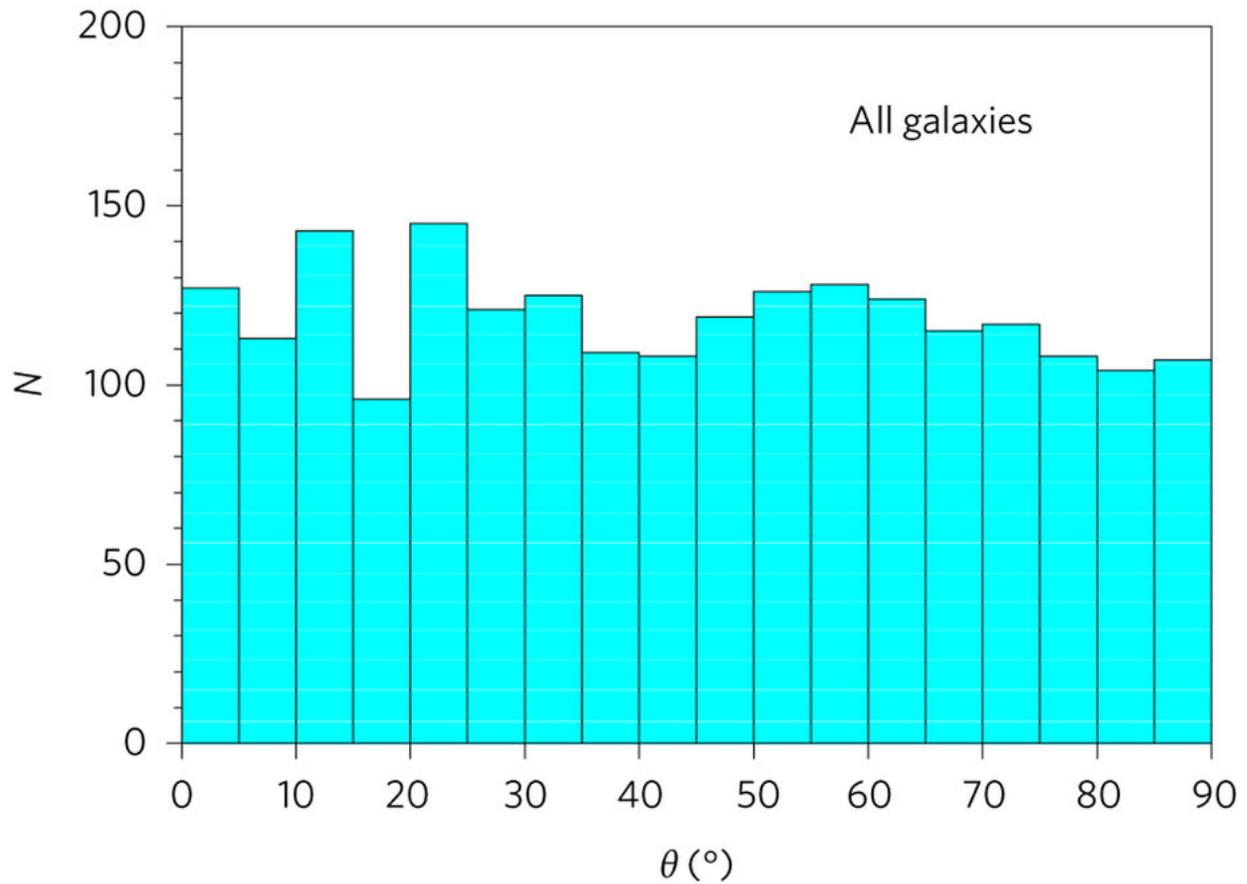

Figure 2. Orientations of 2137 galaxies in CLASH clusters. Here θ is the acute angle between the projected major axis of each galaxy and that of the cluster in which it resides. If the galaxy and cluster axes are perfectly aligned then θ = 0°, while random galaxy orientations will produce a uniform distribution between 0° and 90°. The distribution is consistent with no preferred orientations of cluster galaxies in general.

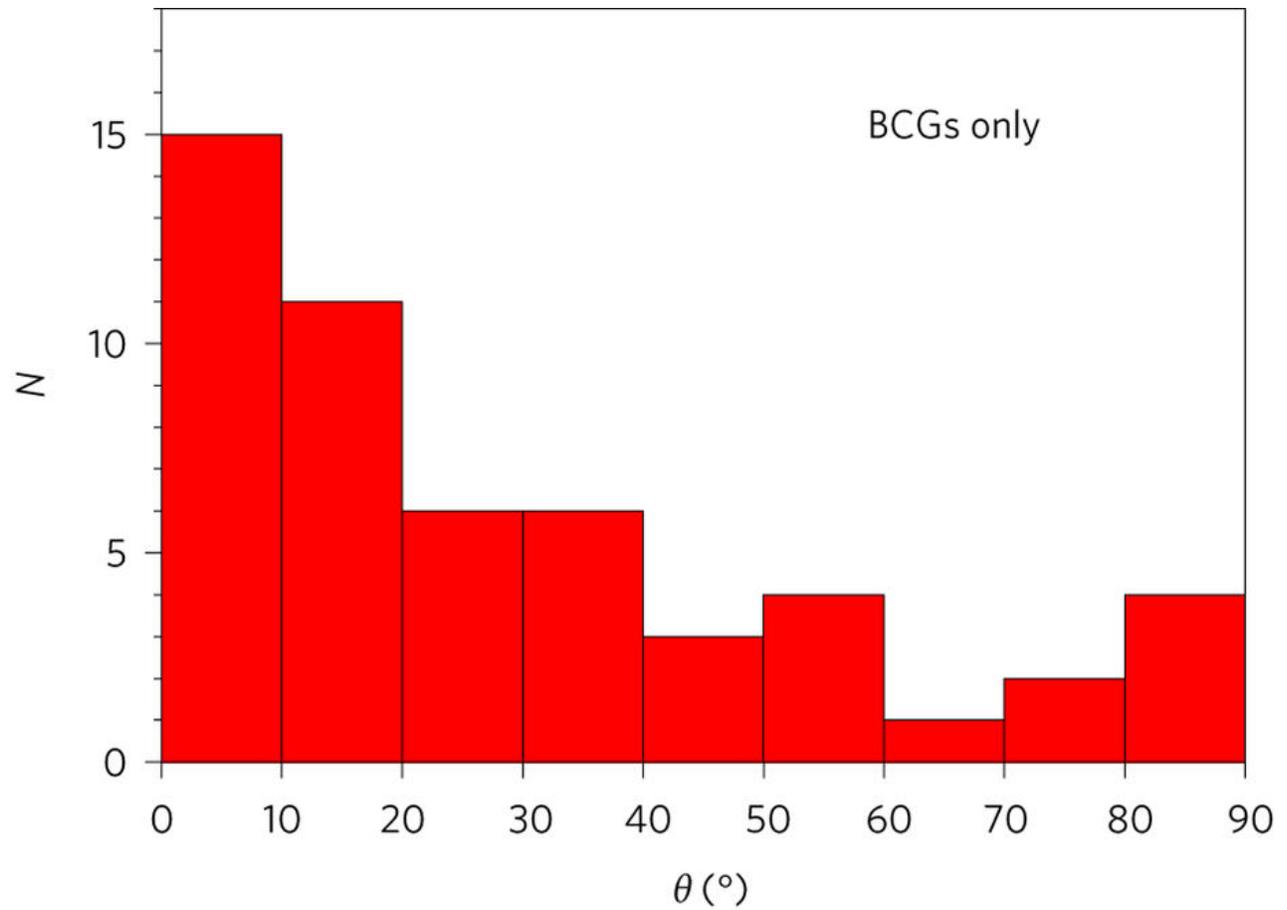

Figure 3. Alignments of brightest cluster galaxies. As in Figure 2, θ is the acute angle between the major axis of the brightest cluster galaxy and that of the cluster in which it resides. A strong tendency for these galaxies to share the same orientation as their host cluster is seen and confirmed statistically.

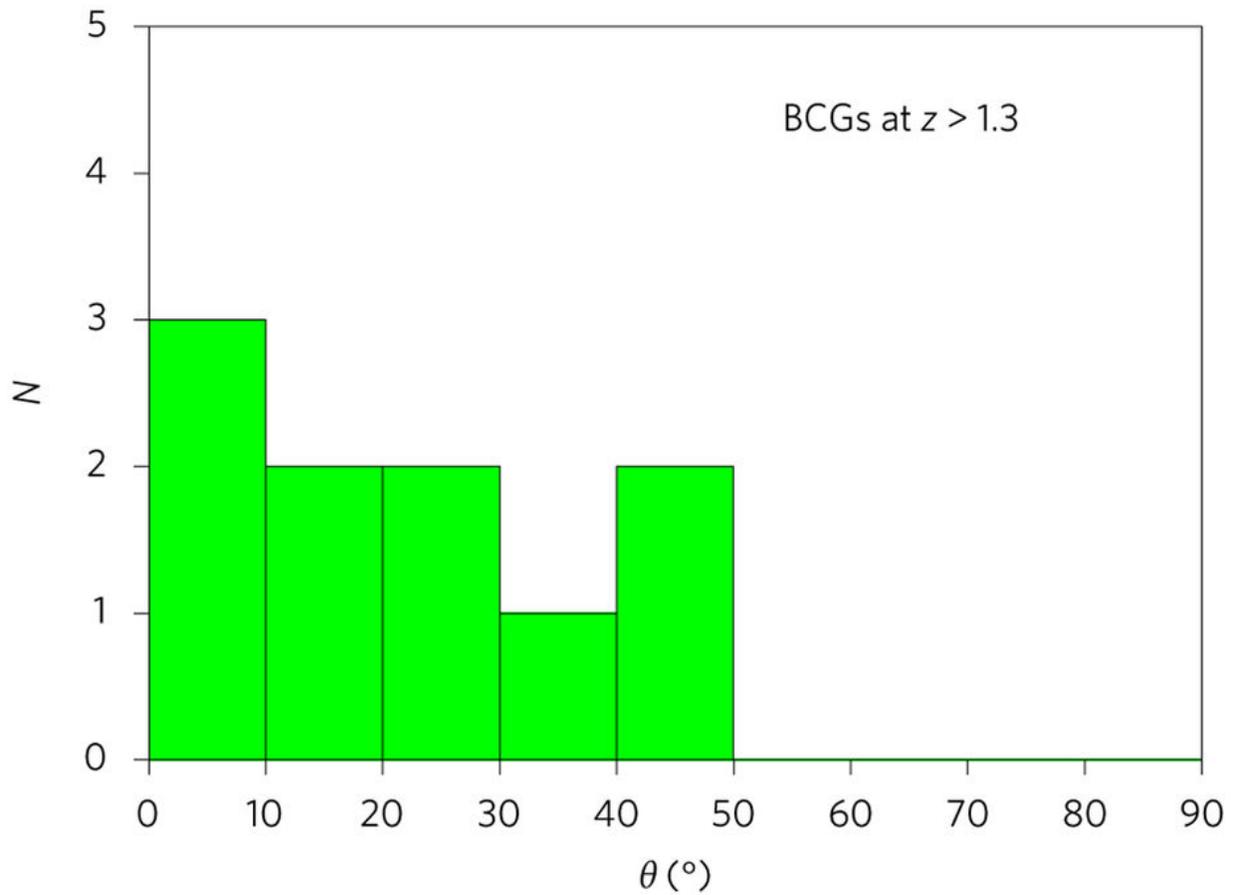

Figure 4. Galaxy alignments at the highest redshifts. The orientations of the brightest member galaxies with respect to their host clusters in the ten clusters at redshifts z > 1.3. A statistically significant tendency for these galaxies and their host clusters to share similar orientations is found even at early epochs.

**Methods**

*Identifying cluster galaxies from the red sequence*

We used the red sequence to select cluster members in the same fashion as previous studies[34,35]. This is based on the tight correlation between color and luminosity for elliptical galaxies, which are the dominant population in clusters. We fit a straight line to the red sequence using the method of minimum absolute deviation to exclude possible outliers. Galaxies within 5σ of the red sequence scatter (±0.25 mag) were assumed to be cluster members. For two clusters in the CLASH sample where spectroscopic information has been published, this yields a fidelity rate great than 95%, although contamination may increase towards higher redshifts and for fainter galaxies[35].

*Position angle measurements and uncertainties*

Cluster position angles were derived from the projected distribution of member galaxies on the plane of the sky. This was done by computing the moments of inertia of the galaxy distribution with respect to the cluster center, which is assumed to coincide with the location of the brightest member galaxy. Position angle uncertainties were estimated by generating 100,000 bootstrap resamples of the member galaxies in each cluster. The 1σ uncertainties are given in Table 1 and are typically 10° to 20°. Clusters whose position angle uncertainty exceeds 25° were culled from the sample.

As an additional check, for CLASH clusters with published mass models derived from gravitational lensing analysis[36], we measured each cluster's principal axis and its orientation by fitting ellipses at a projected distance of 500 kpc from the cluster center using the ellipse task in the Image Reduction and Analysis Facility (IRAF) package. The median difference between these two independent measures of cluster position angles was found to be 11°, consistent with the aforementioned uncertainties.

Individual galaxy orientations were determined using GALFIT[37,38]. Briefly, GALFIT is a data analysis package that fits two-dimensional analytic functions to the surface brightness distribution in extended objects like galaxies. Models are convolved with the image point spread function (PSF) to properly account for the effects of seeing on the derived galaxy structural properties. The PSF is measured empirically from stars in each image, which automatically takes into account the effects of image processing (e.g., drizzling). We note that most of the galaxies in these HST images are substantially larger than the PSF.

GALFIT deblends overlapping galaxies and can fit any number of components simultaneously. Because clusters are dominated by elliptical galaxies, we adopt a single-component Sérsic profile to describe their surface brightness distributions. The Sérsic profile has several adjustable parameters: an index that characterizes the degree of central concentration, luminosity, centroid, axial ratio, half-light radius, position angle and deviations from an ellipsoidal shape. In general, GALFIT is quite robust and insensitive to input parameters, yielding typical uncertainties of less than 5° for the major axis position angles of most galaxies. Further details of how GALFIT works can be found in the original papers.

*Statistical assessment of galaxy alignments*

Determining whether galaxies have preferred orientations with respect to their host cluster requires statistical analysis. We employ three well-established statistical tests to assess the significance of the results shown in Figures 2-4.

The Rao Spacing Test[39] ascertains whether directional data are isotropic. It is a non-parametric test and requires no binning of data. The basic idea is simple: if galaxies have random orientations then the acute angle, θ, between a galaxy's major axis and that of the cluster in which it resides should be uniformly distributed between 0° and 90° and approximately evenly spaced 90°/*N* degrees apart, where *N* is the number of galaxies in the sample. Large deviations from uniform spacing indicate data that are clustered or anti-clustered, as is expected if galaxies have non-random orientations. The test was implemented by sorting the angles θ and then calculating the sum, *U*, of deviations between adjacent values,

$$U = \frac{1}{2} \sum_{i=1}^{N} |T_i - \lambda|$$

where

$$\lambda = \frac{90°}{N}$$

and

$$T_i = \theta_{i+1} - \theta_i \text{ for } i \leq N-1, \quad T_i = (90° - \theta_N) + \theta_1 \text{ for i = N}.$$

The statistical significance of the observed value of *U* for a sample is found by generating a million Monte Carlo realizations in which *N* angles are randomly drawn between 0° and 90° and recording how often the value of *U* found for the random realizations exceeds that of the real data.

Kuiper's *V* statistic[40] provides a second measure of the significance of galaxy alignments. The statistic, which is a refinement of the well-known Kolmogorov-Smirnov test, is defined as

$$V = D_+ + D_-$$

where $D_+$ and $D_-$ are the maximum deviations above and below the cumulative distribution function of the observed values of θ compared to the expected cumulative distribution for random galaxy orientations. The null hypothesis of isotropy is rejected for large values of *V*. Like Rao's Spacing Test, Kuiper's test is non-parametric and free of binning. As before, the statistical significance of *V* values found for the data shown in Figures 2-4 was determined by generating Monte Carlo realizations of *N* randomly chosen values of θ between 0° and 90° and recording the frequency with which *V* for the random samples exceeded that of the real data.

The binomial test provides a third statistical assessment that is useful for small sample sizes. Like a simple coin toss in which there are two possible outcomes, we compare the number of times that θ, which ranges from 0° to 90°, is less than or greater than 45°. The binomial distribution then yields the probability of the observed numbers occurring if galaxy orientations are isotropic,

$$p(x) = \frac{N!}{x!\,(N-x)!}\, 0.5^x\, 0.5^{N-x}$$

where *N* is the total number of angles θ and *x* is the number of times that θ is less than (or greater than) 45°.

**Acknowledgments:** The authors are indebted to Cristóbal Sifón for insightful comments and recommendations that strengthened our results and their presentation. This work is based on observations made with the NASA/ESA *Hubble Space Telescope*, obtained from the data archive at the Space Telescope Science Institute. STScI is operated by the Association of Universities for Research in Astronomy, Inc. under NASA contract NAS 5-26555. M. J. W. thanks The Finnish Centre for Astronomy with ESO (FINCA) and Tuorla Observatory for their support and hospitality during this research. We thank Melissa McIntosh for her assistance. IRAF is distributed by the National Optical Astronomy Observatories, which are operated by the Association of Universities for Research in Astronomy, Inc., under cooperative agreement with the National Science Foundation.

**Author Contributions**: R. D., M. N. B. and S. P. identified red sequence galaxies in the HST images and R.D. measured their major axis orientations. M. J. W. measured cluster position angles and performed statistical analysis of alignments between galaxies and clusters. All authors contributed to interpretation and presentation of the results in this manuscript.

**Data availability:** The data that support the plots within this paper and other findings of this study are available from the corresponding author upon reasonable request.

**Competing interests:** The authors declare no competing financial interests.

**How to cite the published article:** West, M. J., De Propris, R., Bremer, M. N. & Phillipps, S. Ten billion years of brightest cluster galaxy alignments. Nat. Astron. 1, 0157 (2017)